# Delineating hierarchical activity space from high-resolution urban mobility flows

Zhicheng Deng[a,b,1], Zhaoya Gong[a,b,1,*], Jean-Claude Thill[c,d], Elizabeth C. Delmelle[e]

[a]*School of Urban Planning & Design, Peking University Shenzhen Graduate School, Shenzhen, China;* [b]*Key Laboratory of Earth Surface System and Human-Earth Relations of Ministry of Natural Resources of China, Peking University Shenzhen Graduate School, Shenzhen, China;* [c]*Department of Geography and Earth Sciences, University of North Carolina at Charlotte, Charlotte, NC, USA;* [d]*School of Data Science, University of North Carolina at Charlotte, Charlotte, NC, USA;* [e]*Department of City and Regional Planning, University of Pennsylvania, Philadelphia, PA, USA*

[1]These authors contribute equally to this paper

Correspondence: Zhaoya Gong, School of Urban Planning & Design, Peking University Shenzhen Graduate School, Shenzhen, China 518055. Email: z.gong@pku.edu.cn

# Delineating hierarchical activity space from high-resolution urban mobility flows

Current studies on activity space are limited by the conceptualization of absolute physical space that fails to consider the heterogeneity of relational spaces reconstructed from spatial interactions of human movements between locations and falls short in incorporating the inherent hierarchical property of human mobility. Consequently, these approaches cannot faithfully reflect how people interact with urban spaces through travels. From the lens of relational space, this study proposes the new Hierarchical Activity Region Model (HARM) to derive the space and hierarchical properties of activity spaces perceived by various urban groups. We demonstrate the enhanced validity of our model on travel behavior in Manhattan, New York City, before, during, and after Hurricane Sandy on the basis of taxi data. Empirical results show that intra-urban travel retains clear hierarchical organization, even under disruption of a major weather event. Yet, travel undergoes a compression effect in travel hierarchies, characterized by fewer hierarchical levels and enlarged characteristic scales, followed by a rebound. Clustering the derived hierarchies reveals pronounced heterogeneity that stems from differences in population profiles; some groups sustain deeper structures or recover quickly, while others experience a persistent loss of levels. This study provides valuable insights into the functional hierarchies of urban mobility, which could inform more sustainable, resilient and equitable urban planning. The proposed methodological framework is generic for studying human mobility in broader contexts.

**Introduction**

Building safer, more sustainable, and inclusive cities has been a development aspiration worldwide through modern times (Griggs *et al.* 2013, Colglazier 2015, Yue *et al.* 2026). Achieving this goal requires a deep understanding of how people interact with urban spaces through their daily activities. At the core of these interactions lies the concept of activity space, which refers to the set of places an individual visits as part of their daily routines, such as commuting, shopping, leisure, and social interactions (Golledge and Stimson 1997, Cagney *et al.* 2020). Activity space reflects dynamic, behavior-driven perceptions of places and responses to them. It is not simply a static or absolute construct but it is inherently relative, shaped by individual travel decisions and experiences, and external constraints (Desbarats 1983, Miller 1999). As such, activity space serves as a critical lens for understanding and measuring spatial supply-demand dynamics in urban systems.

Activity space has long been a focus for geographers and urban planners (González *et al.* 2008, Wong and Shaw 2011). Early studies used travel surveys and GPS trajectories to model daily movement for relatively small cohorts (Schönfelder and Axhausen 2003, Wan and Lin 2013), whereas the rise of big data has enabled researchers to analyze massive human mobility data and trace activity spaces of urban populations (Wachowicz *et al.* 2016, Chen *et al.* 2018, Li *et al.* 2025). These approaches illuminate different facets of personal activity spaces, including geometric properties such as their spatial extent and their shape (Lee *et al.* 2016, Wang and Li 2016), connectivity through transport networks (Jin *et al.* 2021), the intensity and duration of activities (Zhang *et al.* 2019, Tao *et al.* 2020), and the randomness of activity space (Yuan *et al.* 2012, Yuan and Raubal 2016).

The legacy view on activity space is rooted in a fixed and absolute spatial framework that treats space as a static background or container to human activities (Thill 2011). In contrast, real-world activity spaces are inherently dynamic: they are complex and adaptive, often shifting in response to situational or environmental changes (Dodge *et al.* 2020, Gong *et al.* 2024, Goulias 2024). Therefore, it is befitting to employ a conceptualization of relational space supporting the core principle that people's experience with space and how they use it to engage in certain functions in their daily lives (e.g., human travel behaviors) in turn (re)produces space and shapes how it operates (travel or mobility space) (Couclelis 1992, Harvey 2006, Thill 2011).

As an intersecting consideration, the hierarchical nature of travel behaviors has been well demonstrated in previous studies (Fotheringham 1986, Jeng and Fesenmaier 2002, Jeong *et al.* 2022). For instance, when migrants choose a destination, they typically consider a hierarchy of scales: first evaluating broader areas, such as a state or metropolitan region, before narrowing down to specific districts and neighborhoods. However, research on activity space modeling has paid little attention to the multiscale nature of travel behaviors. Specifically, existing studies typically approach the topic either through the lens of the hierarchies of urban facilities and services on the aggregate (Christaller 1933, Chen *et al.* 2024) or the lens of hierarchical variabilities of individual travel and activity behaviors across geographic scales (Chen *et al.* 2023). These studies, however, are once again grounded in the concept of absolute geographical space with pre-determined place hierarchies as departure point. Hence, they fail to capture that human travel behaviors produce hierarchical structures of activity spaces.

In this study, we propose a model of hierarchical travel spaces to address the shortcomings of existing methods that neglect hierarchies in activity spaces. Dubbed the Hierarchical Activity Region Model (HARM), this model is designed around relational distances and the explicit treatment of spatial heterogeneity in actual behaviors. To estimate the proposed model, we extend an existing framework for individual travel to a localized population cohort with a reasonable assumption of local homogeneity. Specifically, the solution approach of micro-simulation involves a spatio-temporally constrained random walk method to support the generation of micro-level data. Once the hierarchical relational travel spaces are reconstructed, we fully characterize their properties and examine their heterogeneity, explicitly accounting for the hierarchical nature of perceived travel space across population groups. Hence, it closely aligns with the concept of relational space. To demonstrate its validity and effectiveness, we conduct an empirical study in Manhattan, New York City, characterizing the hierarchical activity-space structures revealed by taxi travel, identifying heterogeneous traveler groups, and examining how their structures evolved longitudinally in response to the shock of Hurricane Sandy in October 2012. In the context of this real-world setting, the study investigates three core research questions:

RQ1. How can we operationalize a relational space-time framework to explicitly model the hierarchical and heterogeneous nature of human activity spaces?

RQ2. By applying this framework, what hierarchical patterns of activity space and corresponding nested structures can be quantitatively delineated from the day-to-day travel of urban populations?

RQ3. How do diverse urban residents organize and operate their activity regions to fulfill specific urban functions, and what sets apart these patterns across population groups?

The next section of this paper provides a review of the literature on measuring human mobility in relational space, as well as on the concept of activity space. The following section introduces the Hierarchical Activity Region Model, including its theoretical grounding, estimation procedure, and the clustering method to investigate heterogeneity of activity regions. The fourth section describes the study area and data processing. The fifth section presents the empirical case study and results, and discusses the findings. The final section concludes with the overall contributions of the paper to literature in spatial modeling, urban analytics, and spatial flow analysis.

**Literature Review**

*Spatial Organization of Human Mobility in Relational Space*

As a crucial aspect of spatial interaction, human mobility has always been a focal point in geographic research (Ullman 1980, Roy and Thill 2003, Kwan and Schwanen 2016, Oshan 2021). It refers to the movement of human beings (both individuals and groups) in space and time, reflecting a dynamic interaction between people and their environment. Characterizing mobility requires the conceptualization of space, and within spatial science, discussions on the concepts of space are roughly divided into two primary categories: absolute space and relative space (e.g., Couclelis, 1999; Massey, 1999; Ullman, 1974), as illustrated in Figure 1.

Figure 1. Characterizing human movements in absolute space (a) and in relative space (b).

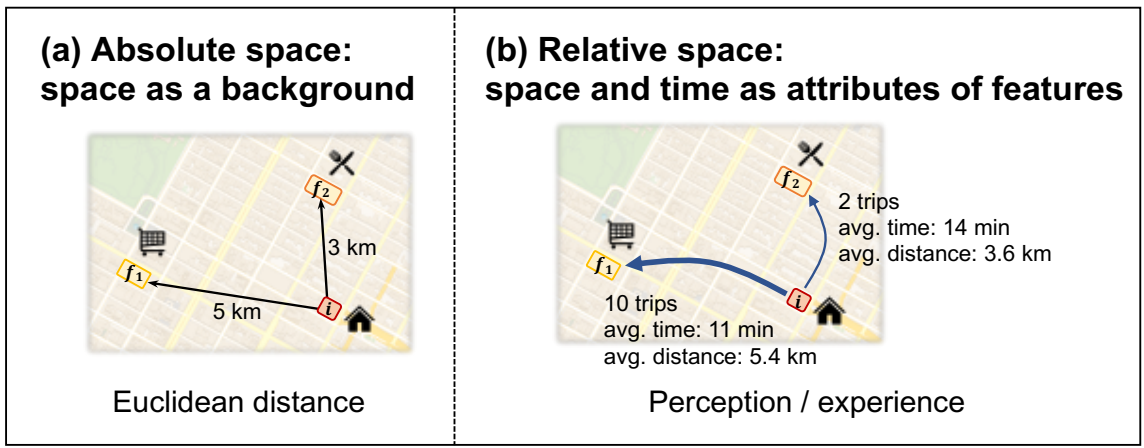

In the concept of absolute space, space is objective, independent of the entities being investigated, and consistently measurable (Curry 1996). It acts as a container and serves as the background for all features, with non-dynamic space-time structures. Movement is often represented by two-dimensional physical locations and by the Euclidean distances between them, which act as measures of travel cost (Huff and Jenks 1968, Plane 1984). These travel costs are treated as static, unaffected by time or external contexts.

Relative space, by contrast, considers space and time as attributes of features, allowing for better analysis and expression of the semantic content of spatial objects (Thill 2011, Shaw and Sui 2020). Individuals perceive and experience space differently, shaping their unique understandings of the world. For example, with a given distribution of amenities, individuals may vary in their choices of local destinations (Malekzadeh *et al.* 2024), and each of these travel behaviors carries distinct temporal and spatial characteristics—such as variations in frequency, duration, and distance.

Building on the foundations of relative space, researchers have advanced the concept of relational space, in which space and time are constituted and defined by processes and human behavior (Harvey 2006). Within this framework, spatial structures emerge from relational processes—how people move, interact, and assign meaning to places. While recent studies have begun to describe urban space as shaped by patterns of human mobility (Li *et al.* 2021), they often overlook the heterogeneity in how different population groups construct and experience these relational geographies through their travel behaviors. This gap motivates a more differentiated examination of mobility-generated spatial structures across spatial contexts.

### *Delineating and Modeling Activity Space*

Activity space is a fundamental concept in geography and urban studies, referring to the set of locations individuals—or even entire groups—routinely visit during their daily activities (Golledge and Stimson 1997, Schönfelder and Axhausen 2003, Yuan and Xu 2022). Although scholars employ a variety of related terms, they converge on the idea of a personal or collective *spatial footprint* that spans home, workplace, shopping, leisure, and other routine destinations (Brown and Moore 1970, Hägerstrand 1970,

Horton and Reynolds 1971). It thus serves as a lens through which researchers can examine spatial accessibility, social interaction, and exposure to environmental influences (Kwan 1998, Kar *et al.* 2023, 2024).

Researchers have developed various methods to empirically delineate activity spaces. The first category is geometric metrics, which summarize overall size and shape of activity space—examples include the radius of gyration (González *et al.* 2008) and movement eccentricity (Yuan *et al.* 2012). Other studies adopt a network-based perspective, focusing on location nodes and their connections, as reflected in metrics such as distance travelled (Tao *et al.* 2020, Wang and Yuan 2021, Li *et al.* 2025). Additionally, some approaches characterize activity spaces by considering the number and duration of visited locations (Wang and Li 2016, Zhang *et al.* 2019), as well as the uncertainty of visitation patterns (e.g., entropy) (Song *et al.* 2010, Yuan and Raubal 2016).

These approaches, however, overlook the inherently hierarchical nature of travel decisions. For example, when going shopping, individuals often make a series of choices within multiple levels—first choosing whether to stay within their district, then narrowing down to specific streets, and finally selecting a particular store. This process generates distinct segments of travel distance at each level. Recent work has introduced the so-called container model to capture characteristic spatial scales of mobility (Alessandretti *et al.* 2020), but it remains rooted in an absolute, flat conception of space. As such, it does not explicitly model people's diverse spatial perceptions and behaviors, nor does it construct a travel-centered, hierarchical activity space. Furthermore, it does not explain how these nested scales emerge from travel behaviors driven by the pursuit of different urban functions.

**Methodology**

*Hierarchical Activity Region Model*

This section introduces a place-based container model, dubbed Hierarchical Activity Region Model (HARM), to endogenize hierarchical structures of relational spaces reconstructed from human activities and travels. Our model is place-based as we assume that people living in the same place (here we term it the *home location*) tend to exhibit similar travel behaviors. Conversely, groups of people residing in different places with varying attributes have varied travel experiences and perceive space differently. Hence their activity spaces may exhibit varying geographical and hierarchical structures, reflecting the heterogeneity of activity spaces of travel behaviors. By incorporating relational distances that better represent travel behaviors into HARM, we can derive hierarchical travel spaces for groups of people from different home locations. This approach extends the original container model (Alessandretti *et al.* 2020), which focuses exclusively on individual mobility and is limited to the Euclidean distance as the metric for space reconstruction.

Let us consider the relational activity space of a group of people from the same home location and depict it by a hierarchical structure *H*, which hierarchically subdivides the relational space encompassing all places accessed by the group. This division is organized through a tree structure of nested *activity regions* with *L* levels, where each activity region is constituted by a set of places; hence, the activity region at the root level refers to the entire activity space while each activity region at the leaf level is an individual place (Figure 2). A set of child activity regions at the lower level coalesces to form a parent activity region at the next higher level, provided that the relational distances between child activity regions are within a certain critical proximity. This critical proximity reflects the spatial scale of activity regions at a certain level, which characterize various boundaries of human settlements (e.g., neighborhood, district, and city). Consequently, $H$ embodies a multiscale spatial division, characterized by a series of discrete spatial scales $\{s_l, l \in L\}$ (referred to as the *size of level*).

Figure 2. Diagram of the container model. People move between locations (black dots) within nested activity regions (polygons). The transition between two locations is described as a two-stage decision process. Adapted from Figure 1c in Alessandretti *et al.* (2020).

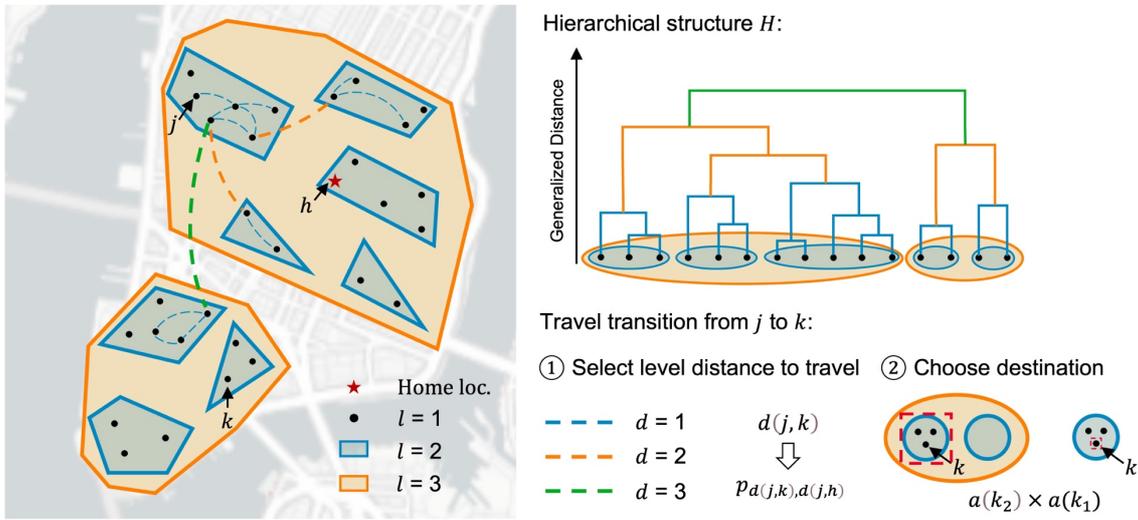

We first define the distance $D_{ij}^l$ between any two activity regions $AR_i^l$ and $AR_j^l$ at level $l$ as the longest distance between places within the two regions:

$$D_{ij}^l = \max_{u \in AR_i^l, v \in AR_j^l} Distance(u, v), \qquad (1)$$

where $Distance(\cdot)$ can be Euclidean or relational (travel time or distance) distances. Then, the size $c(\cdot)$ of an activity region $AR_i^l$ is defined as the longest distance between any two child activity regions within $AR_i^l$:

$$c(AR_i^l) = \max_{AR_j^{l-1}, AR_k^{l-1} \in AR_i^l} D_{jk}^{l-1} \qquad (2)$$

Thus, $s_l$ the size of level $l$ can be defined as the size of the largest activity region among $n_l$ activity regions at level $l$:

$$s_l = \max_{i \in n_l} c(AR_i^l) \tag{3}$$

Consequently, for two activity regions $AR_i^{l-1}$ and $AR_j^{l-1}$ at level $l-1$ that belong to a same parent activity region at level $l$, the distance between them must satisfy the following condition:

$$s_{l-1} < D_{ij}^{l-1} \leq s_l. \tag{4}$$

where $D_{ij}^{l-1} \leq s_l$ is straightforwardly derived from equations (1)-(3), while $s_{l-1} < D_{ij}^{l-1}$ must be valid otherwise $AR_i^{l-1}$ and $AR_j^{l-1}$ would be merged into a single activity region at level $l-1$.

In the original container model, Euclidean distances are used as the distance measure in equation (1), which is not a proper metric for reconstructing relational spaces of travel activities. Because the separation between two locations is not solely determined by their physical distance (such as Euclidean distance) but rather by the costs of spatial interactions that occur as people travel. The latter can be more faithfully depicted in terms of travel distance and travel time. To better represent this type of relational spaces, HARM incorporates travel distance and travel time as measures of relational distances, capturing human perceptions, experiences, and activity interactions within these spaces (Gould 1991).

The model also introduces the concept of *level distance*, which is defined as the highest level where the activity regions differ when traveling from place $j$ to $k$ while traversing a sequence of nested activity regions at different levels. Level distance measures the number of spatial scales that need to be crossed for a trip from $j$ to $k$;

hence it takes integer values up to $L$: $d(j,k) = 1, 2, \ldots, L$. Then, the transition probability of a travel transition from place $j$ to $k$, $P_{H,a,p}(j \to k)$, is determined by a two-stage decision process (see Figure 2): the probability people choose to overcome a certain level distance and the probability that a particular destination is selected given the chosen level distance. HARM follows this modeling approach:

$$P_{H,a,p}(j \to k) = p_{d(j,k),d(j,h)} \frac{a(k_{d(j,k)})}{1-a(j_{d(j,k)})} \prod_{l=1}^{d(j,k)-1} a(k_l), \qquad (5)$$

where $p_{d(j,k),d(j,h)}$ represents the conditional probability of transitioning to a level distance $d(j,k)$ given the current level distance from home $d(j,h)$. The term $\prod_{l \leq d(j,k)} a(k_l)$ represents the probability of selecting activity regions that include location $k$ at level distances less than $d(j,k)$. The attractiveness of any activity region $k_l$ at level $l$ is denoted $a(k_l)$. It represents the probability that this activity region will be selected within its parent region at the next higher level. The attractiveness of all activity regions at the same level sums to 1: $\sum_{k_l \in k_{l+1}} a(k_l) = 1$. Equation (5) models location choices over a spatial hierarchy reconstructed by nested activity regions and is related to nested logit models for discrete choices (Williams 1977, McFadden 1978).

*Estimation of HARM*

This section describes the methods to estimate the proposed HARM for a certain home location and the construction of key variables to that effect. Given the purpose of this model, inputs consist of a set of trips undertaken by people from the featured home location over a certain time period. To incorporate this into the original container model, which was developed for trips of an individual, we assume that 1) the group from the featured home location exhibits behavioral homogeneity in terms of movement patterns, which is supported by the theory of time geography and further empirical research on human mobility (Kwan 1998, González *et al.* 2008, Zhang and Thill 2017); 2) the set of visited locations depends on the population at the home location—that is, the larger the population at the location, the more trips generated, and the more locations visited.

In our case study, we use hourly origin-destination (OD) flows to construct the input variable, given that individual trip data are not available. OD flows represent the aggregates of individual trips between locations. Due to this property, once people leave their home locations, the locations they subsequently visit become untraceable in the flow data. To address this issue, we develop a spatio-temporally constrained random walk method taking advantage of the hourly OD flows to approximate those untraceable trips. Specifically, two key spatio-temporal constraints are considered in this micro-simulation method. First, we consider the home location as a fundamental anchor for mobility behavior, consistent with anchor-point theory (Couclelis *et al.* 1987). Under this constraint, people are assumed to leave their home location within a certain periodicity that aligns with empirical observations, such as the 24-hour periodicity of taxi travel behavior in Manhattan described by Zhu and Guo (2017). Second, after leaving their home location, people will follow a path to return to the starting point within a confined time frame (e.g., 24 hours), reflecting the cyclical nature of human mobility.

The algorithm for the spatio-temporally constrained random walks with hourly OD flow data is illustrated in Figure 3. We account for the temporal distribution of in- and out-flows of a home location by using a multinomial distribution to simulate one's departure time $T_d$ and return time $T_h$ with the constraint $T_d \leq T_h$. Then, one chooses the next location following a random walk process weighted by the spatio-temporal flows from origins to destinations; once one visits a new location, a new departure time $T_d^{'}$ needs to be determined according to the temporal distribution of out-flows from the current location. This process iterates until the time reaches $T_h$ when one is required to return to the home location immediately. During this process, the time and distance of trips between each visited location, including those involving the home location, are recorded; if no trip occurs, it is recorded as missing. Practically, this process needs to be repeated many times to ensure sufficient trips sampled for a home location and repeating times are weighted by the population size of the home location according to our assumption 2. By incorporating these constraints, this method not only enables the inference of visited locations for a group of people from a home location but also ensures that the generated mobility patterns remain consistent with the observed OD flow distributions.

Figure 3. The workflow of spatio-temporally constrained random walk algorithm for inferring visited locations of a group of people.

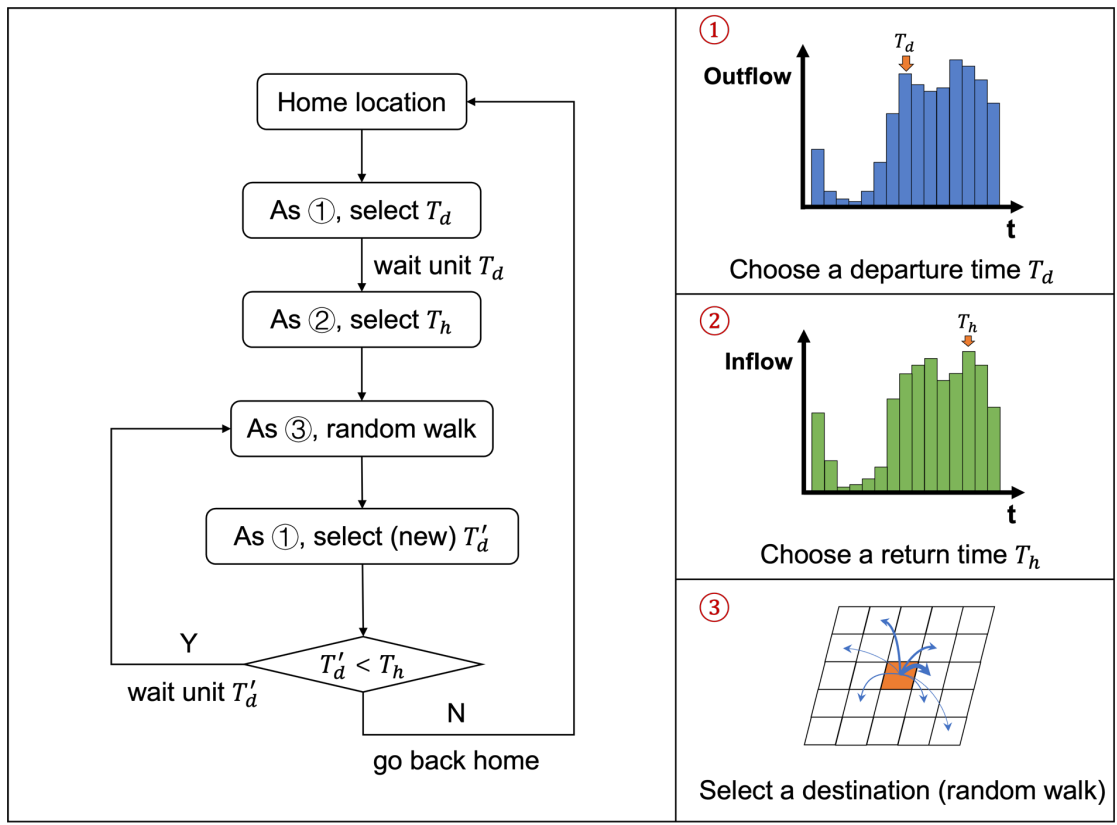

After constructing the data variables for people from a specific home location, we can fit a container model to them and obtain the hierarchical structure $H$, attractiveness $a$, and transition probability $p$. Distances between different locations are used as input for a complete linkage algorithm (Murtagh and Contreras 2012) to generate an initial structure $H^0$. Specifically, every location is merged with another according to the shortest distance, and the bottom-up merging process continues hierarchically until forming the initial tree $H^0$. To estimate the final $H$ and related model parameters, we follow the maximum likelihood method adopted in the original paper. The likelihood function is given by:

$$L(H, a, p|T) = \prod_{i=0}^{n_T-1} P_{H,a,p}\big(k(i-1) \to k(i)\big). \tag{6}$$

*Examining Heterogeneity of Hierarchical Activity Spaces via Clustering*

We examine the spatial heterogeneity of hierarchical activity spaces through clustering home locations based on summary metrics derived from HARM. For each home location $i$, we assemble a feature vector that captures the hierarchy: $x_i = [L_i, \overline{D}_i, ...]$, where $L_i$ is the number of levels in the estimated hierarchy, and $\overline{D}_i$ is the mean level distance of trips computed from the observed trips under the estimated hierarchical structure. Because level distance counts the number of spatial scales crossed, both $L_i$ and $\overline{D}_i$ are measured on the same dimensionless scale $[1, L_i]$; in particular, the maximum level distance for a given hierarchy equals $L_i$.

Clustering is performed with the *k*-means algorithm with the squared Euclidean objective, $\min_{\{C_k\},\{\mu_k\}} \sum_{k=1}^{K} \sum_{i \in C_k} \|x_i - \mu_k\|_2^2$. We use k-means++ initialization to improve convergence and multiple random restarts to avoid poor local minima. The algorithm iterates between assigning observations to the nearest centroid and updating centroids until the relative decrease in the objective falls below a small tolerance or a maximum iteration limit is reached.

To determine the number of clusters $K$, we combine a goodness-of-fit criterion with a separation–cohesion criterion. First, we compute the within–cluster sum of squares ('inertia') for candidate values. We then locate the elbow on the inertia curve using the Kneedle algorithm (Satopaa *et al.* 2011), as implemented in the kneed package, which identifies the point of maximum curvature after normalizing the relation between K and inertia. The selected knee marks the value beyond which additional clusters yield diminishing returns in fit.

As a complementary check, we evaluate the average Silhouette over all observations for each $K$. For an observation $i$, let $a(i)$ be the mean distance from $i$ to all other points in its assigned cluster (cohesion), and let $b(i)$ be the minimum, across all other clusters, of the mean distance from $i$ to points in that cluster (separation). The Silhouette value is defined as $s(i) = \frac{b(i)-a(i)}{max\{a(i),b(i)\}}$, with larger values indicating better separation and tighter cohesion. We regard as optimal a $K$ that coincides with the Kneedle-identified elbow on the inertia curve and attains a local maximum of the average Silhouette with a relatively large value, indicating a favorable balance between model parsimony and cluster quality. Building on the resulting partition, we can further investigate the heterogeneity of hierarchical activity regions formed by different groups of people, and how people operate their mobility space to fulfil their daily functions differently.

**Case Study**

*Study Area*

Manhattan, located at the heart of New York City, is one of the most iconic and vibrant boroughs in the United States. Covering approximately 59.1 square kilometers, Manhattan is densely populated and serves as a major economic, cultural, and commercial hub. It is renowned for its dense gridiron street layout, a defining feature of its urban geography. The borough is organized into a numbered street system running east to west and avenues running north to south. While its grid structure aligns with rectilinear distance metrics in an absolute space, this static representation fails to capture the dynamic realities of movement in such a bustling and congested environment. As a result, people's travel experiences and perceptions of space in Manhattan are far from uniform. The same physical distance can represent vastly different travel costs or impedances depending on traffic conditions and individual behaviors. Consequently, a measurement framework rooted in a relative space context is required, which can provide a more accurate representation of distances and travel costs in this distinctive urban space.

*Data*

Three datasets were assembled: a taxi trip dataset, a census dataset, and a land use dataset. The taxi trip dataset was sourced from the New York City (NYC) Taxi and Limousine Commission (TLC). This dataset serves as a benchmark that has been widely utilized and is readily accessible. It specifically records all yellow[1] taxi trips and covers three weeks encompassing the Hurricane Sandy event, from October 21, 2012, to November 10, 2012. We further divided this period into three phases (pre-event, during-event, and post-event), reflecting documented shifts in Manhattan taxi flows at the time of the hurricane (Bian *et al.* 2019). Using these phases, this empirical study will compare the hierarchical structures of human mobility within relational spaces before, during, and after the hurricane's impact.

---

[1] The green taxi service was inaugurated in August 2013.

The dataset records individual, point-to-point taxi trips (including information like pick-up location, drop-off location, time, and distance, see Table 1). Several preprocessing steps were undertaken on these data, including matching the spatial extent and cleaning outliers. A total of 7.11 million trip records were selected.

Table 1. Extract of NYC taxi data

|   | Pickup_time | Pickup_x | Pickup_y | Dropoff_time | Dropoff_x | Dropoff_y | Trip time (s) | Trip distance (mile) |
|---|---|---|---|---|---|---|---|---|
| 1 | 2012-10-21 00:00:00 | -73.983406 | 40.738926 | 2012-10-21 00:05:00 | -73.978165 | 40.729488 | 300 | 0.98 |
| 2 | 2012-10-21 00:00:00 | -73.982895 | 40.739292 | 2012-10-21 00:09:00 | -73.955116 | 40.777473 | 540 | 3.16 |
| 3 | 2012-10-21 00:00:07 | -73.985954 | 40.738541 | 2012-10-21 00:05:51 | -73.992455 | 40.730598 | 343 | 1.00 |
| 4 | 2012-10-21 00:00:10 | -73.983253 | 40.738739 | 2012-10-21 00:03:58 | -73.973755 | 40.752251 | 228 | 1.00 |
| 5 | 2012-10-21 00:02:00 | -73.986336 | 40.740181 | 2012-10-21 00:14:00 | -73.981110 | 40.725368 | 720 | 1.53 |

Although the cleaned dataset consists of individual taxi trips, each record is merely a single OD pair and cannot form the necessary input of HARM directly (which requires a sequence of visited locations for a group sharing the same home tract). To obtain a group-level input that preserves overall flow patterns while respecting privacy, we aggregated trips to the census-tract level. Figure 4 illustrates this aggregation for October 21 using FlowMapper (Koylu *et al.* 2023).

Figure 4. Spatial distribution of the 200 most frequent taxi trips on October 21$^{st}$ (aggregated to census tracts).

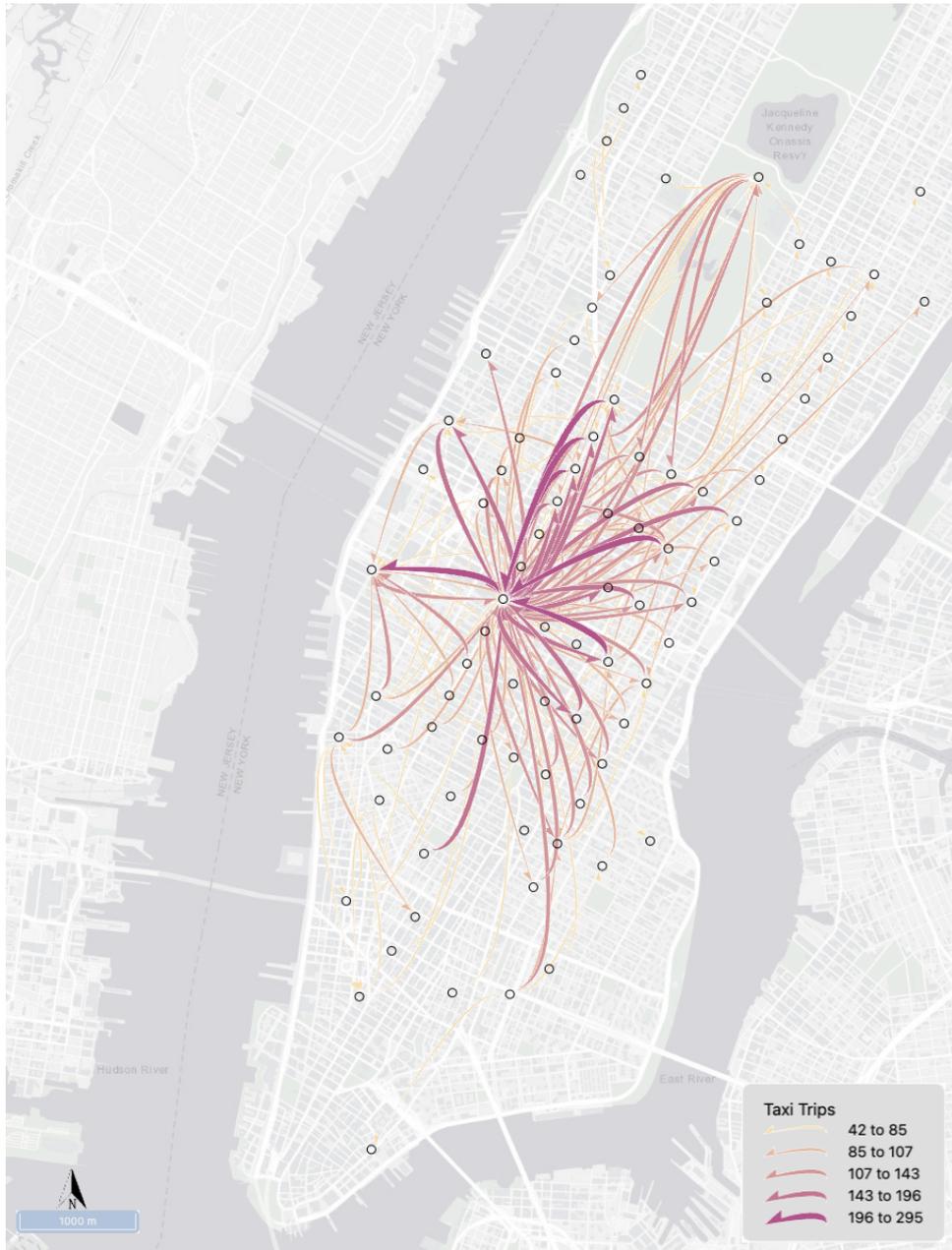

To ensure adequate sampling, we retained only tracts with at least 100 departures or arrivals per week and a resident population of 1,000 or more, yielding 236 tracts for analysis (Figure 5). Trips were further aggregated into one-hour intervals. On this spatio-temporal granularity, we employed a spatio-temporally constrained random walk approach (details provided in the Methodology) to infer visited locations of people from different home tracts. Validation tests (Appendix A) show that the reconstructed visit patterns closely match observed mobility across all three hurricane phases.

Figure 5. Study area and its population.

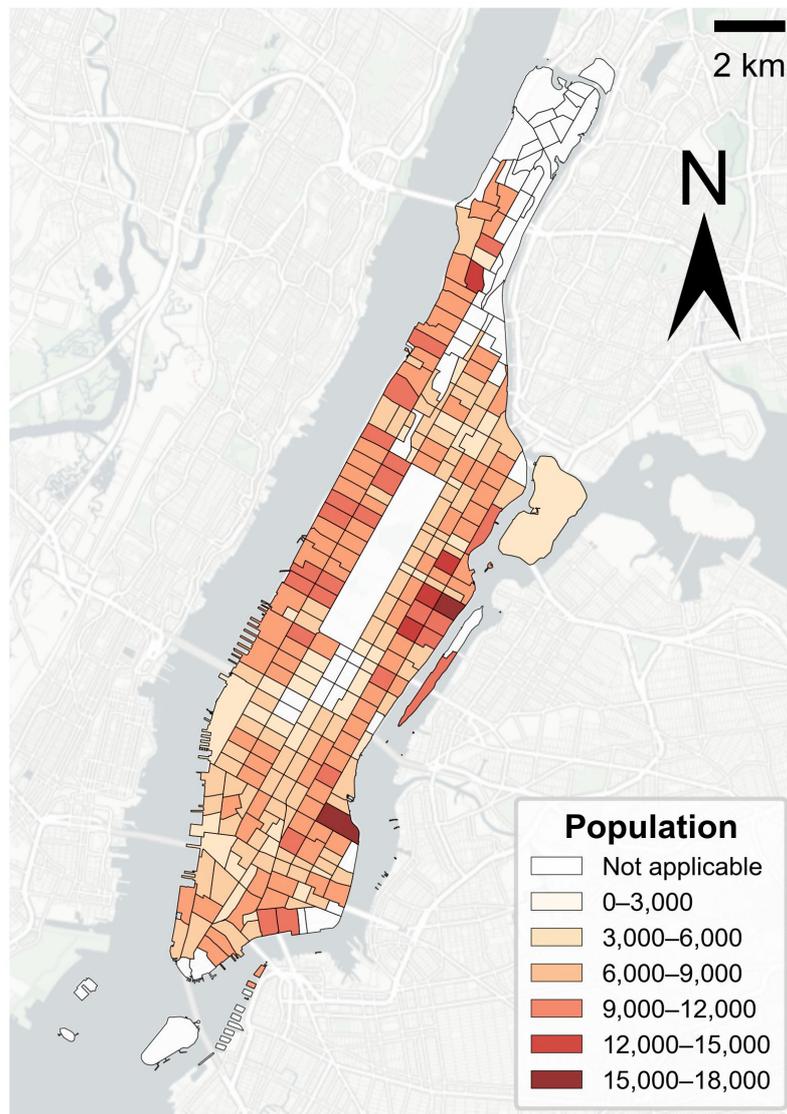

The land-use dataset was selected to approximate how people organize activity spaces to obtain specific functions. We employed the NYC Department of City Planning's MapPLUTO 12v2, which classifies parcels into eleven categories (Figure 6): 01–03 residential types, 04 mixed residential–commercial, 05 commercial/office, 06 industrial/manufacturing, 07 transportation/utility, 08 public facilities/institutions, 09 open space/outdoor recreation, 10 parking, 11 vacant. We followed this classification without modification. To link mobility and the built environment, taxi drop-off points were associated with land use by assigning each point the land-use type using the nearest-neighbor matching. This yields a land-use label for every observed destination, enabling comparisons of the relationship between hierarchical distance and function types across different phases.

Figure 6. Land-use map of Manhattan, with colors showing the category.

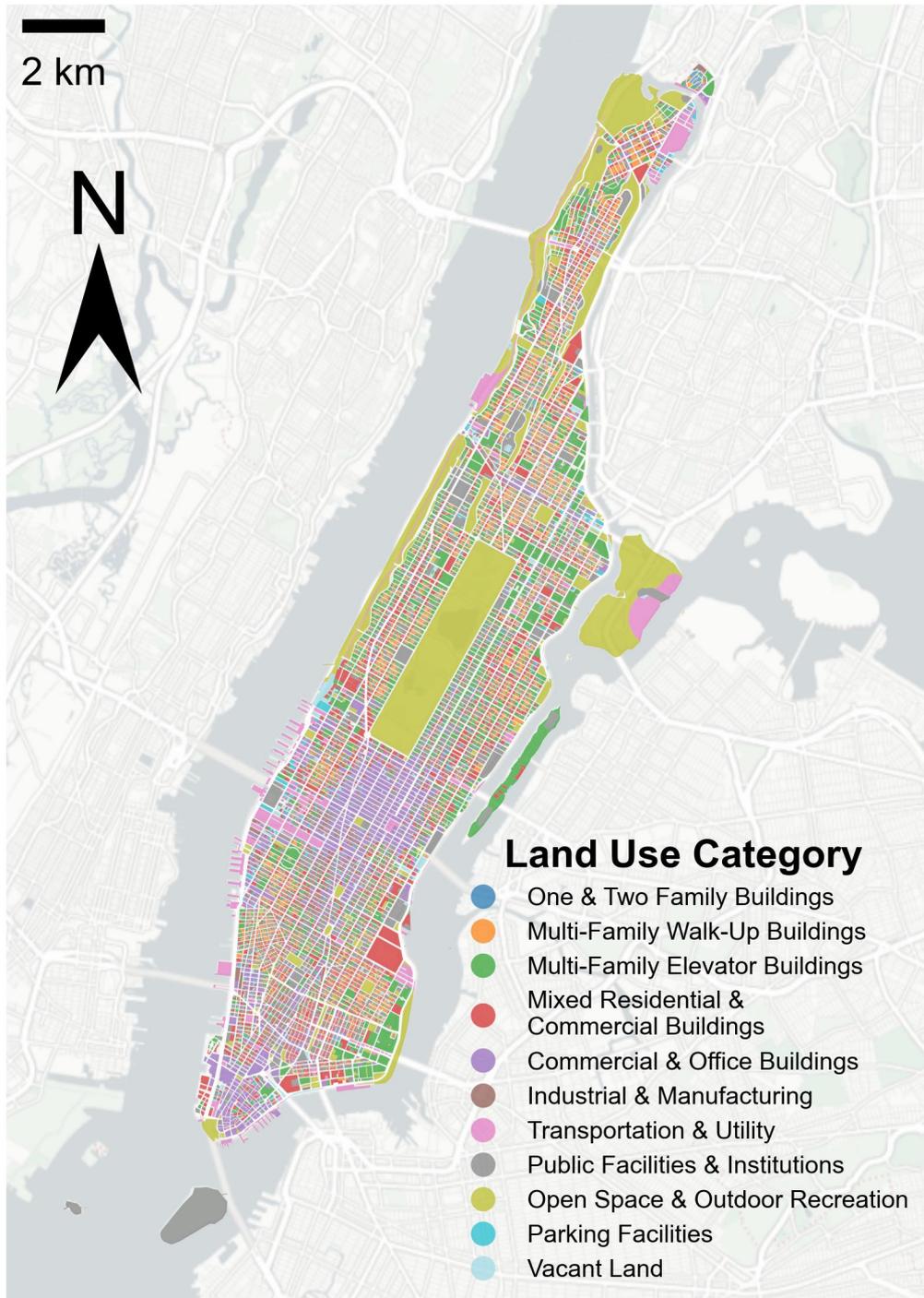

**Results and Discussions**

*Hierarchical Activity Regions in Relational Versus Euclidean Space*

First, we estimate the HARM with Euclidean distance as the spatial metric (referred to as the HARM-Euclidean), in turn taking each census tract as a home location. In the hierarchical structure of any arbitrary location (Figure 7a), each level exhibits a spatial partitioning that is encapsulated within the levels above it. Here, we illustrate this with the case of a branch of the hierarchical structure generated for a home location in the Civic Center (census tract # 001300) during the Hurricane event. A dendrogram is used to demonstrate the process of merging individual locations to form higher-level clusters via the complete linkage algorithm. Dendrogram cuts correspond to the hierarchical partitioning of individual locations, and different dashed lines indicate the corresponding spatial characteristic scales (also called size of levels). For instance, individual locations are represented at level 1, while at level 2, activity regions might correspond to neighborhoods, and at a higher level (e.g., level 5), they could represent districts. Within each level, different activity regions signify different locations, neighborhoods, or districts. Locations within a certain proximity are grouped into the same neighborhood, while certain neighborhoods are grouped together in a single district partition, and so on.

Figure 7. An example of the hierarchical structure generated using Euclidean distance (a) and relational distances (b) as measures of space for a home location in the Civic Center (census tract # 001300).

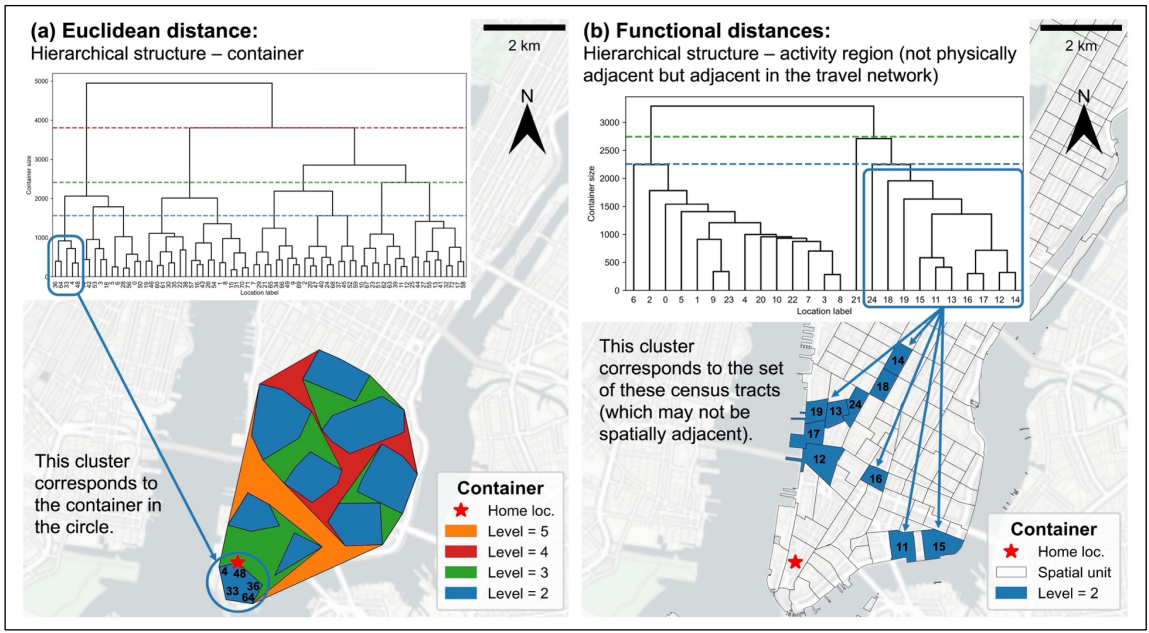

Next, to gain a more accurate characterization of the hierarchical travel space, we use the HARM to incorporate relational distances instead of the Euclidean distance (denoted as HARM-relational). Drastically different structures come to light with this specification. Figure 7b depicts the hierarchical structure generated for the same home location as in Figure 7a during the event, but based on travel time. In this scenario, a nested hierarchical structure is still evident through the dendrogram. Most notably though, we find that the relational distance creates activity regions that may include locations that are not physically adjacent (blue tracts in Figure 7b).

Broadening to the scope of the entire study area, a distinctive property of HARM is that it can detect and represent complexity in travel behaviors. Specifically, a larger number of levels in the estimated HARM is indicative of a broader and more diverse range of spatial choices for activities, while fewer levels suggest a more concentrated selection of travel spaces. With this in mind, we can underscore the differences between the HARM-Euclidean and the HARM-relational in terms of level numbers within the respective hierarchies. By the same token, we can analyze how the hierarchical organization of taxi travel was affected by Hurricane Sandy (see Figure 8).

It is evident that during the event, the number of levels decreased compared to the pre-event phase, leading to what we term 'scale compression'. This is manifested by a decrease in the frequency of higher levels and an increase in the frequency of lower levels (since the total frequency is fixed, which corresponds to the number of spatial units – census tracts). In terms of actual travel behavior, extreme weather conditions lead people to undertake essential trips only, which can cause certain characteristic spatial scales to 'disappear.' For example, the travel hierarchy generated for a home location in Chinatown (census tract # 001600) saw a decrease in the number of levels from 10 to 5 during the hurricane. People from this home location localized their travel behaviors, curtailing long-distance trips that cross multiple areas, resulting in the disappearance of higher levels in the generated hierarchical structure. After the event, the number of levels rose again, even above the pre-event number.

Also, compared with the original Euclidean space (characterized by HARM-Euclidean), the hierarchies defined in relational travel space (whether distance or time based) are deeper and more finely articulated across all phases (Figure 8b and c). They partition the city into behaviorally meaningful activity regions and therefore capture the hierarchical organization of actual travel choices more accurately. Notably, the hierarchy based on travel time (Figure 8c) exhibits the clearest pattern: during the event, the distribution shifts sharply toward lower levels with minimal overlap across phases, while post event, it rebounds and extends further into higher levels. The steeper slopes around the disruption and the longer high-level tails after the event indicate a stronger compression-then-rebound dynamic than in panels (a) or (b).

Figure 8. Distribution of number of levels before, during and after the event across the hierarchical structures generated using Euclidean distance and several alternative relational distances.

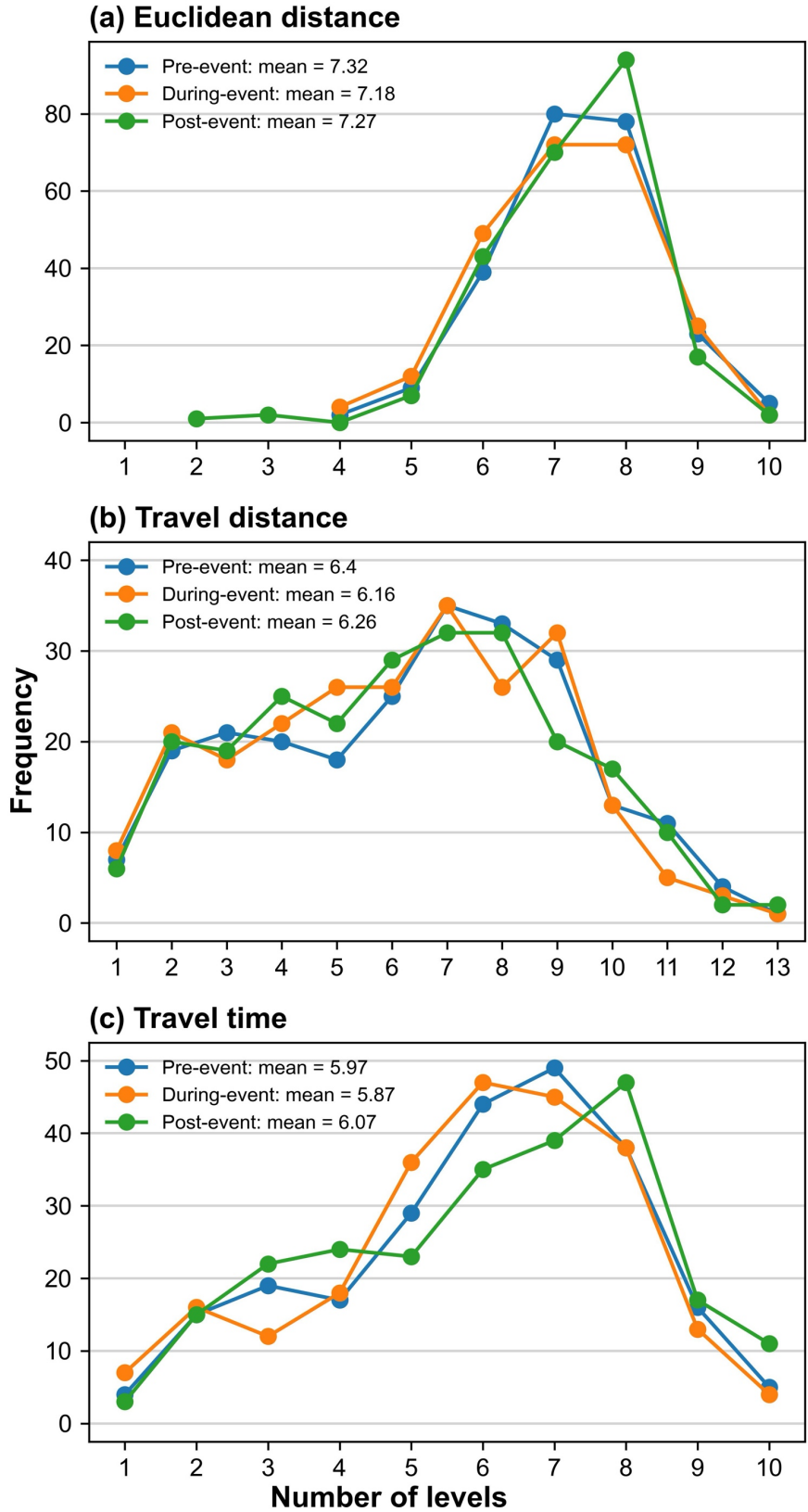

As previously mentioned, HARM allows us to identify different characteristic spatial scales of travel. Hence, we can characterize the size of each level within the corresponding space. Taking the hierarchy constructed based on travel time as an example, we examine the distribution of sizes of level under different phases of the disruption, as shown in Figure 9 (see Appendix B for a comparison of results under different spatial measures). We can see that the sizes of level show a significant increase during the event compared to the pre-event phase, followed by a decrease post-event. For example, the mean size of level 6 increased from 48.6 minutes to 51.8 minutes, and came back to 46.9 minutes after the hurricane. In line with the previous findings that the number of levels decreases, we observe an overall compression of spatial hierarchy denoted here by fewer levels being encapsulated within spaces of the same size. This phenomenon indicates that during extreme weather events, people prioritize essential trips, simplifying their travel patterns. After the hurricane, people might engage in what can be described as 'revenge travel,' leading to a more complex hierarchical structure in their mobility patterns.

Figure 9. Distribution of sizes of level before, during and after the hurricane based on the hierarchical activity regions constructed using travel time.

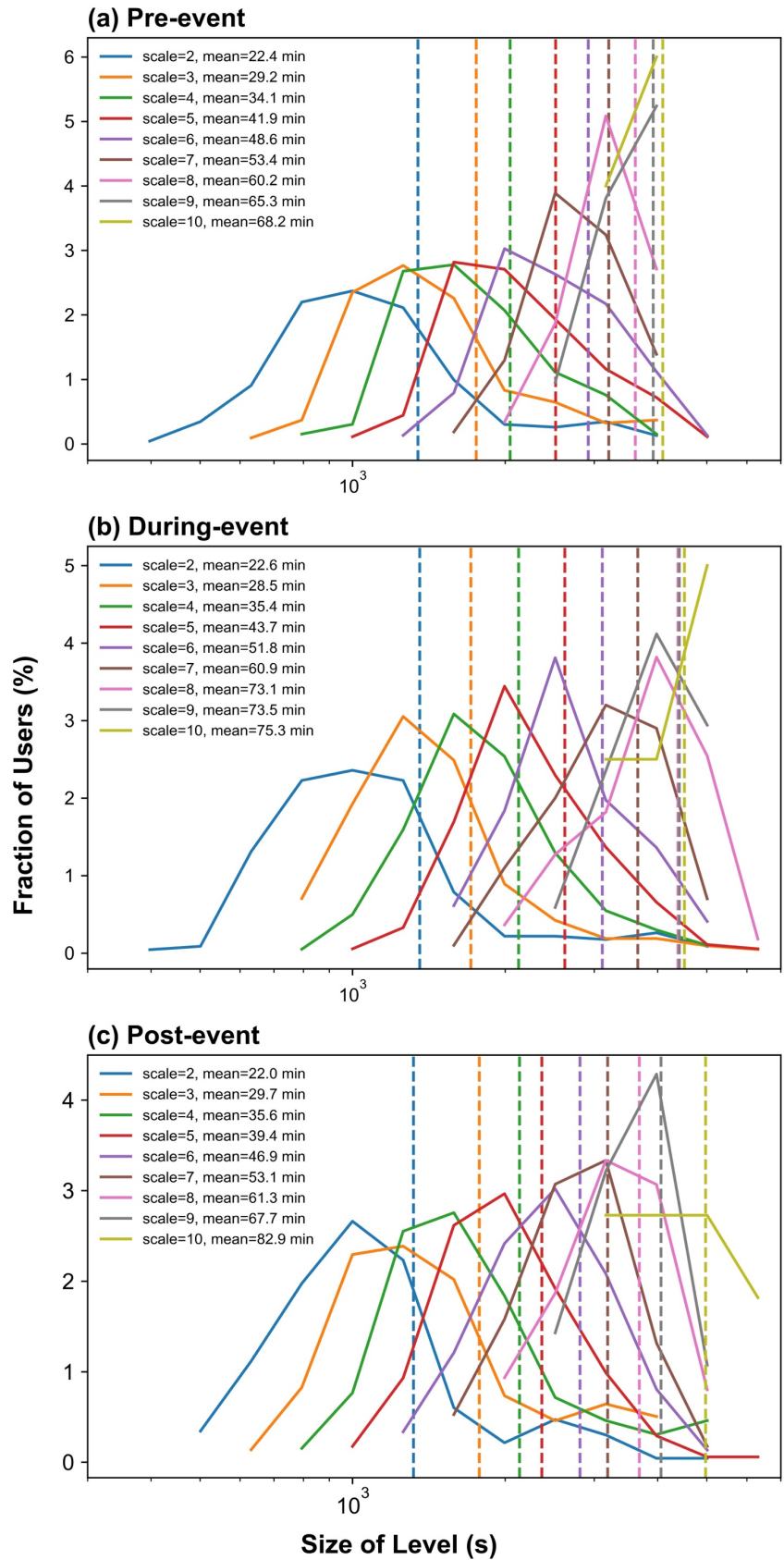

Another key property of HARM is the level distance, which represents the largest characteristic scale traversed by a single trip within the estimated hierarchy. Trips within the same neighborhood typically exhibit small level distances, while cross-district movements are associated with larger values. For any home location, the empirical distribution of level distances defines a probability density function that captures the hierarchical distance effect. Each home location has a probability density for selecting different level distances, with the sum of these densities equaling one. Figure 10 illustrates these distributions constructed for individual home locations based on travel time, showing both the aggregated distribution across all trips and the density of level distances computed separately for individual spatial units (see Appendix B for comparisons of level distance frequencies under alternative spatial measures).

Figure 10. Distributions of level distance density for the hierarchy constructed based on travel time before, during and after the event. (a) Overall probability density of level distance aggregated across all trips. (b) Density across home locations.

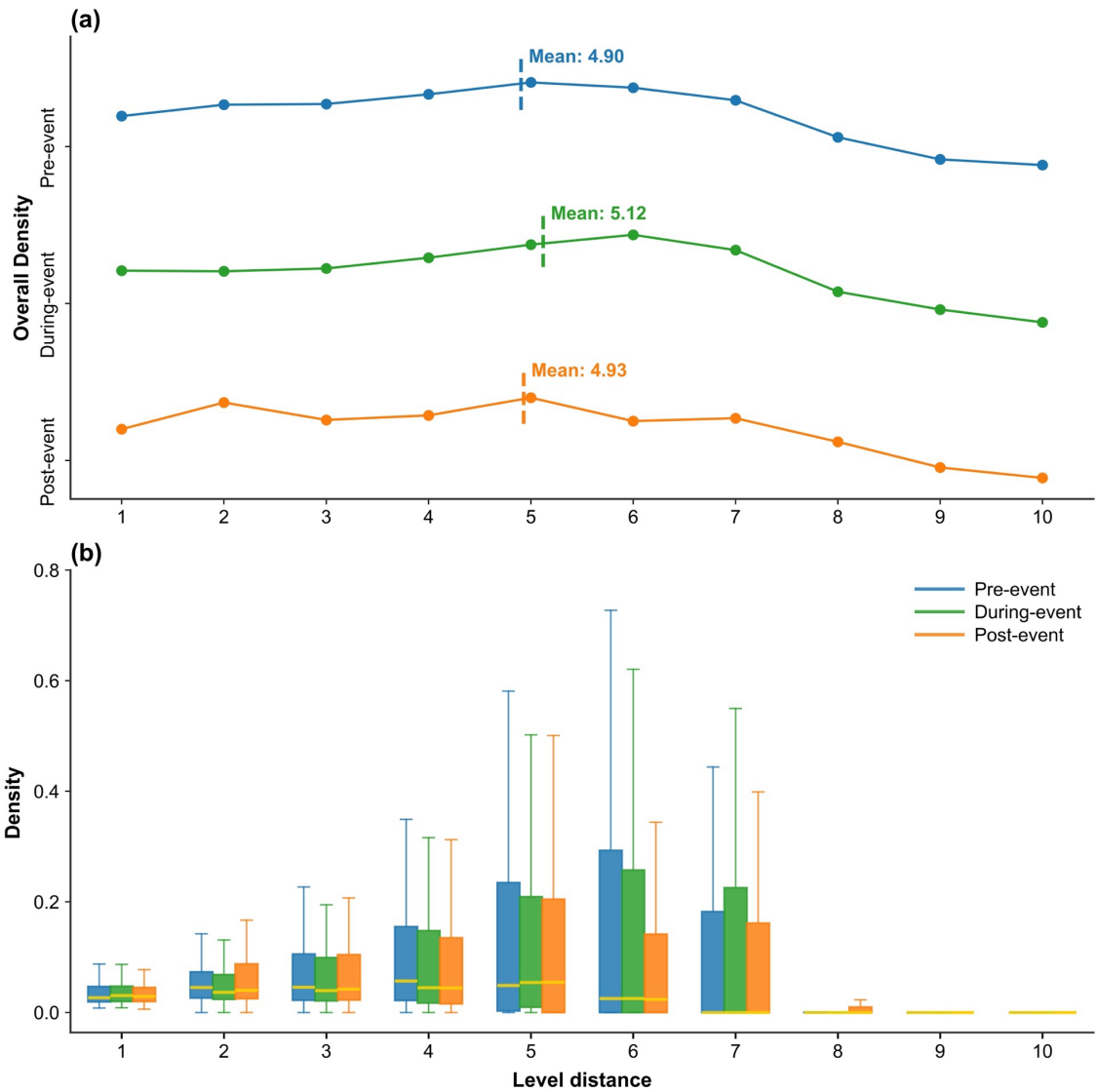

Two patterns stand out. First, pronounced heterogeneity is observed across home locations. Some origins concentrate their movements within a limited range of levels, while others span multiple scales. This is due to the varying complexity and number of levels in the hierarchical structures constructed for different home locations, leading to a broad and differentiated distribution of level distances. Second, the overall mean level distance across all home locations increased during the hurricane event (from 4.90 to 5.12) and subsequently returned to the pre-event levels (4.93) afterward. Although the overall hierarchy exhibited 'scale compression' in terms of fewer levels and larger activity regions, this did not imply a uniform contraction across all trips. Instead, it indicated a selective contraction: localized, short-range trips diminished more sharply during the disruption (e.g., level distance ⩽ 4), while longer-distance, cross-level trips (such as level distances of 6 or 7) were relatively preserved and became more concentrated. Consequently, the overall mean value shifted toward higher values. After the event, as local trips gradually resumed, the mean level distance declined toward its pre-event level.

Previous studies have mostly focused on the hierarchical structure of human mobility using Euclidean distance measures which are biased towards absolute space under standard conditions (Alessandretti *et al.* 2020). In this study, we enhance this research by measuring travel behavior in terms of actual travel distance and travel time, allowing us to examine hierarchical mobility structures in the context of people's lived experiences such as a disruptive weather event like Hurricane Sandy. In summary, results demonstrate well-defined hierarchical patterns of intra-urban activity spaces and they perdured despite the devastating impact of Hurricane Sandy with some notable adjustments. Additional analyses, such as the relationship between the number of levels and the residential population, as well as comparisons across different distance measures, further quantify the effects of the hurricane on the hierarchical structures of human mobility (see Appendix B for more details).

Furthermore, our findings indicate that relational distances, especially travel time, may offer a more accurate, more precise and richer way to capture the intricacies of travel behavior, particularly the compression effect observed during external shocks. Based on hierarchies constructed from travel time (HARM-time), we found that the overall hierarchical character of activity space weakened during the hurricane, while cross-level, longer-range movements were relatively more pronounced. Also, the travel hierarchy exhibits significant heterogeneity across home locations, which reflects distinct activity spaces and behavioral responses.

### *Heterogeneity of Hierarchical Activity Regions Reflecting Distinct Socio-demographic Profiles*

To examine how hierarchical travel structures vary across Manhattan, we leverage six metrics derived from HARM-time to examine heterogeneity: the pre-event number of levels and pre-event average level distance, the during-event change in number of levels (during − pre) and change in average level distance (during − pre), and the post-event recovery in number of levels (post − pre) and recovery in average level distance (post − pre). We are not reporting on the size of level because it is strongly correlated with the number of levels and changes in the same direction, with both jointly indicating 'scale compression'. These variables are standardized and then input into a K-means procedure to partition home locations in the study area into clusters with distinct hierarchical travel structures. Figure 11 presents the diagnostic curves for selecting the number of clusters. In the inertia plot, an elbow appears at $k$=6, indicating a favorable balance between model complexity and within-cluster dispersion. In the silhouette plot, $k$=6 attains a local maximum, providing convergent evidence for this choice. Therefore, we set $k$=6 to delineate six heterogeneous types of hierarchical activity regions in subsequent analyses.

Figure 11. Diagnostics for selecting the number of clusters based on HARM-time: (a) inertia and (b) silhouette score.

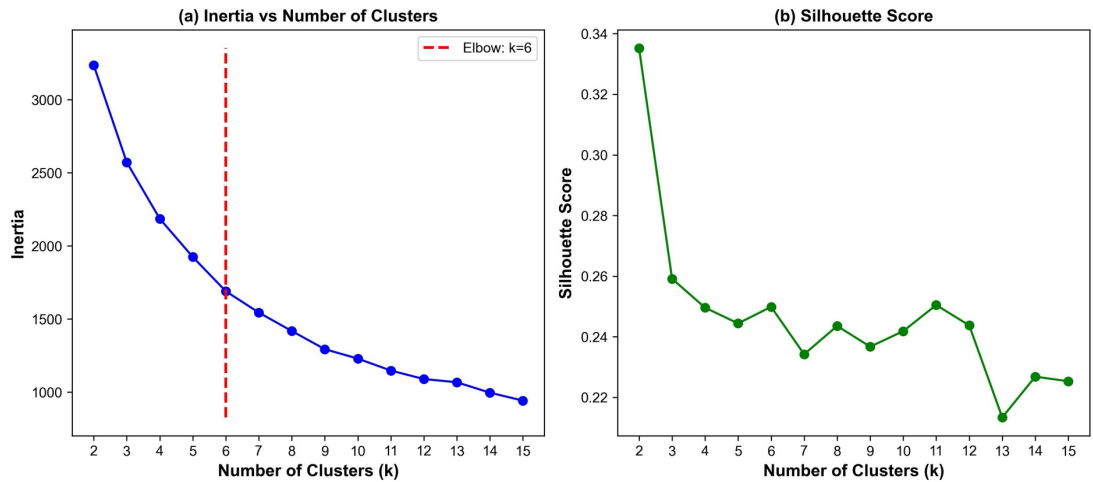

Figure 12 maps the spatial distribution of clusters and shows their value distributions. We obtain relatively balanced results between clusters in terms of the number of home locations (observations) they contain, while significant heterogeneity exists between different clusters. Among them, as seen in Figure 12b, Cluster 1 is the activity-space category with the most travel hierarchies; it experienced the largest contraction during the event and did not fully recover subsequently; Cluster 2 has the fewest levels during the normal stage, but its number of levels rose markedly during and after the event; Cluster 3 has a small and stable number of levels, with little change in either level counts or level distance across phases. Clusters 4-6 have a moderate number of levels, but the change patterns are different: Cluster 4 simplified during the event with fewer cross-level movements and then largely recovered afterward; Cluster 5 became more complex during the event, with increased cross-level movements, and subsequently returned to baseline; Cluster 6's structure simplified during the event, reduced cross-level travel, and showed no clear recovery in the post-event phase.

Figure 12. Spatial layout of the six clusters (a) and distributions of their feature values (b) based on HARM-time.

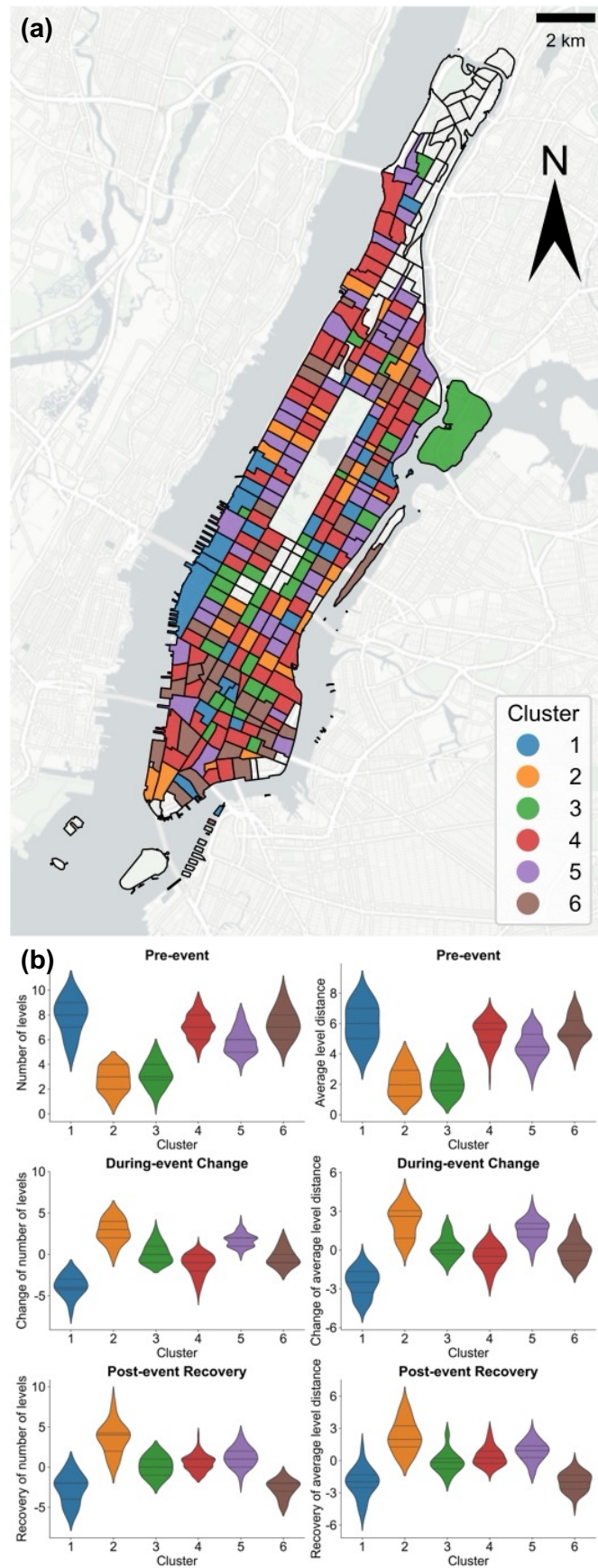

To further the understanding of activity-space properties of the six clusters, table 2 decomposes Figure 12(b) by reporting how average level distance to specific land-use categories changes within each cluster. The reported change rate is defined as the relative change in average level distance during the hurricane period compared to the pre-event baseline, calculated as (during − pre) / pre. Within each cluster, patterns across different functions remain heterogeneous. In Cluster 1, where level distance decreased, *Transportation & Utility* contracted the least, while *One & Two Family Buildings* experienced the largest declines. Clusters 2 and 3 display the same minimum value patterns, suggesting that taxi trips linked to low-density residential uses (often associated with higher-income households) weakened most during the hurricane. By contrast, in Clusters 4–6, the relatively service-oriented categories represented by *Public Facilities & Institutions* and *Open Space & Outdoor Recreation*, performed the weakest, consistently with their nonessential nature under disruption. Functions related to system maintenance and production, such as *Transportation & Utility* (Cluster 1) and *Industrial & Manufacturing* (Clusters 2 and 5), either increased or declined the least, indicating essential travel that is sustained even under adverse conditions.

Table 2. Change rates (%) in average level distance to different land use by cluster. Bold value indicates the largest value in each cluster, while underlined value denotes the smallest.

| Land Use Type \ Cluster | 1 | 2 | 3 | 4 | 5 | 6 |
|---|---|---|---|---|---|---|
| One & Two Family Buildings | <u>-40.97</u> | 89.85 | <u>-2.03</u> | -6.03 | 31.44 | -1.12 |
| Multi-Family Walk-Up Buildings | -39.45 | 92.08 | -0.08 | -7.18 | 32.03 | 0.30 |
| Multi-Family Elevator Buildings | -38.84 | 92.07 | 2.29 | -8.12 | 30.88 | -1.05 |
| Mixed Residential & Commercial Buildings | -37.61 | 95.60 | 3.35 | -7.76 | 32.45 | -0.42 |
| Commercial & Office Buildings | -38.37 | 101.39 | 3.89 | -7.35 | 31.81 | -0.51 |
| Industrial & Manufacturing | -37.54 | **110.59** | 6.00 | -5.28 | **38.12** | -0.61 |
| Transportation & Utility | **-34.72** | 100.41 | 6.06 | -8.44 | 35.72 | -1.29 |
| Public Facilities & Institutions | -39.24 | 91.98 | 5.60 | <u>-8.48</u> | 32.33 | 0.82 |
| Open Space & Outdoor Recreation | -38.84 | 94.88 | 2.67 | -7.06 | <u>30.33</u> | <u>-2.07</u> |
| Parking Facilities | -40.14 | 110.15 | **11.83** | **-5.24** | 32.38 | -1.06 |
| Vacant Land | -35.76 | 100.34 | 2.70 | -8.37 | 31.51 | **2.10** |

Figure 13 summarizes the distributions of socio-demographic attributes across clusters. Clusters 1 and 2 have higher female shares than the others, whereas Cluster 3 has the highest male share. In terms of race, Clusters 1 and 2 are predominantly White, Cluster 3 is predominantly Asian/Pacific Islander, and Clusters 4–6 have higher shares of Black or other racial groups, with Cluster 5 the highest. By income, Clusters 1 and 2 have the highest median household income and also a higher median age, while Cluster 5 represents the lower-income group. These patterns help explain why groups make different travel decisions and, in turn, form distinct hierarchical activity regions.

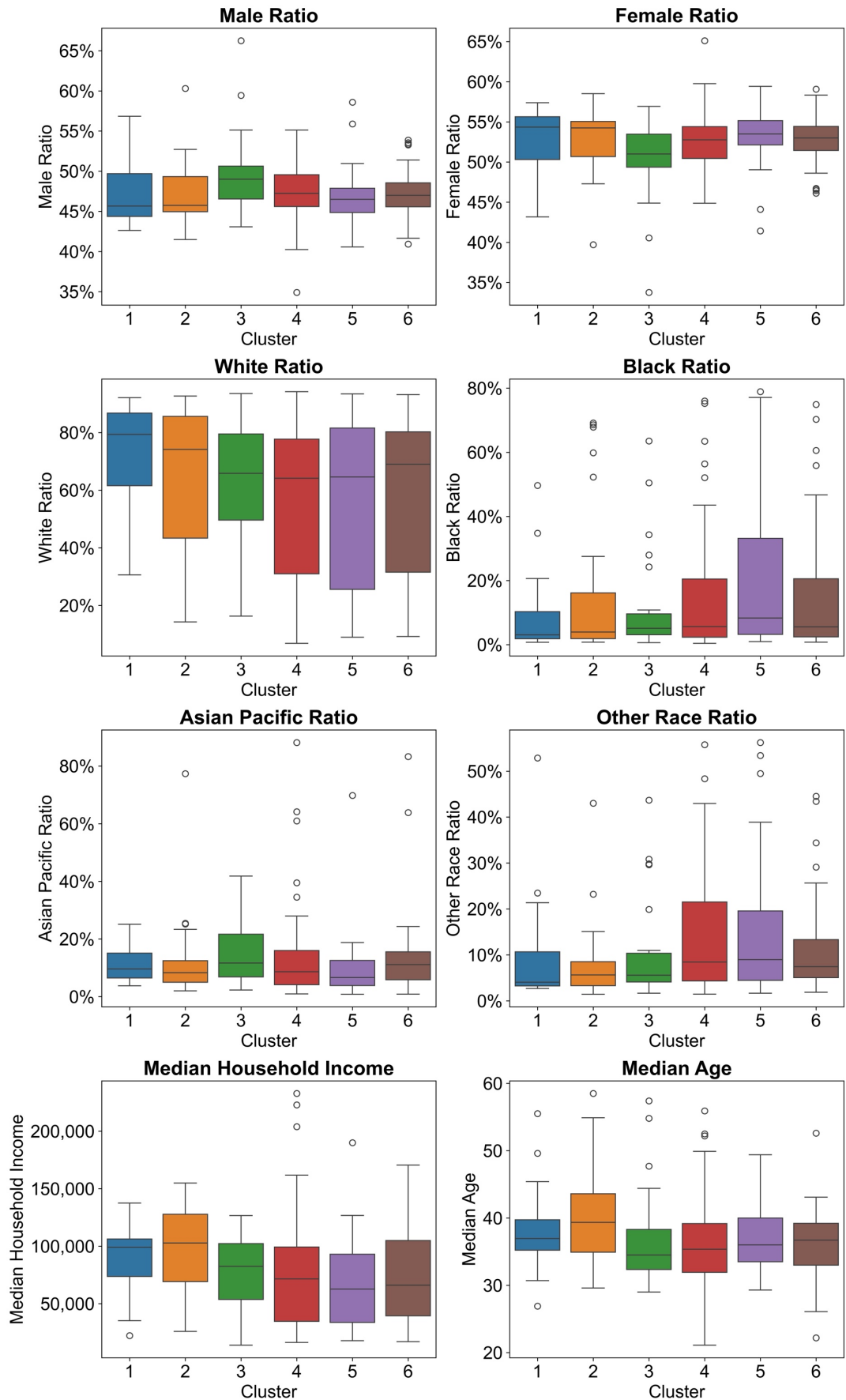

Figure 13. Distributions of socio-demographic attributes across clusters.

With these results on hand, we can document pronounced heterogeneity in intra-city hierarchical mobility patterns of urban residents. Whereas Alessandretti *et al.* (2020) considered socio-demographic differences primarily at larger scales, our analysis operates at the intra-city scale and demonstrates how groups organize hierarchical activity spaces according to local socio-spatial conditions and how these structures relate to urban functions. By integrating socio-economic profiles of residents in these areas, we can interpret the mechanisms behind their formation and their responses to the hurricane. This is discussed in more detail below.

Cluster 1 exhibits the deepest pre-event hierarchies, the largest contraction during the storm, and incomplete recovery. These tracts concentrate in mid- and lower-Manhattan areas dominated by higher-income white residents, with comparatively younger populations and a higher share of females. Although these residents are frequent taxi users in ordinary times (New York Metropolitan Transportation Council and North Jersey Transportation Planning Authority 2014), they chose to proactively avoid travel and curtail visits to residential land-uses under the hurricane.

Cluster 2 has the fewest levels at the normal stage, yet it shows a marked rise in levels during and after the event. These tracts tend to be at spatial edges, with older populations, a slightly higher male share, and a white majority. Despite simple hierarchies in normal times, work commuting needs, such as trips to *Industrial and Manufacturing* or *Commercial and Office Building* areas, we observe sustained and even expanded cross-area movements during the hurricane, with greater reliance on taxis.

Cluster 3 is the most stable profile, with little change across phases. These tracts are characterized by higher Asian shares, younger populations, and middle incomes, often in compact, self-contained southeastern neighborhoods such as Chinatown. This stability may stem from a tightly nested live-work structure within the closely knit community. It is evidenced by the number of levels and average level distances that in these areas are among the lowest during the normal stage (second only to Cluster 2), indicating a relatively simple activity-space hierarchy. As daily activities are concentrated within a small number of spatial levels, reliance on cross-hierarchical travel is limited. Therefore, when Hurricane Sandy disrupted the broader urban mobility system, these activity spaces required minimal reorganization and exhibited small structural changes.

Cluster 4 covers medium-density, mixed-use areas with diverse racial composition and predominantly younger residents. This group experienced a temporary decline in mobility levels during the hurricane but recovered rapidly, indicating that despite facing initial contraction pressures, they possess strong adaptive capacity. This further demonstrates the mobility resilience of certain population groups.

Cluster 5 is concentrated in Harlem and parts of the Midtown periphery, with higher Black shares and lower incomes, and is dominated by minority groups. Here the hierarchy deepened during the storm, implying sustained or intensified cross-district travel embedded in urban maintenance and service work (e.g., trips to *Industrial and Manufacturing*). Despite limited socioeconomic resources, the post-event increases in both the number of levels and the average level distance were largely retained, suggesting a long-lasting reconfiguration of these essential urban functions.

Cluster 6 shows simplification in hierarchical activity spaces during the hurricane with no clear recovery afterward. These tracts are located near the edges of Harlem and in affected parts of the Financial District, with more mixed racial composition, lower incomes, and older populations. Lacking both recovery resources and strong compulsory travel demands, these areas exhibit persistent loss of hierarchical levels and warrant priority attention in providing neighborhood and social services.

Overall, the clustering results show pronounced intra-urban heterogeneity in how hierarchical activity spaces are reorganized under external shocks. Different population groups exhibit different patterns of compression, persistence, and recovery. These findings underscore that hierarchical structures of urban mobility respond non-uniformly to external shocks across space and over time, and they are closely related to the local socio-spatial structures and essential urban functions.

**Conclusions**

In this study, we proposed HARM to explicitly address the hierarchical and heterogeneous nature of activity spaces in relational space. We operationalized this framework by extending an existing framework of individual travel to accommodate group-level mobility and introducing a spatio-temporally constrained random walk method to support the construction of HARM models (RQ1). Examining this framework on travel behavior in Manhattan during Hurricane Sandy, we not only evaluated different mobility patterns and their hierarchical structures but also captured how these structures adapt to external shocks (RQ2). Results show that intra-urban travel retains clear hierarchical organization under disruption yet undergoes a compression effect in travel hierarchies, characterized by fewer hierarchical levels and enlarged characteristic scales, followed by a rebound.

We further examined heterogeneity in hierarchical activity spaces across different population groups and revealed that structural differences in activity spaces are closely related to socio-demographic profiles, including income, age, race, and gender (RQ3). We find that, among other social groups, higher-income residents tend to have hierarchical activity spaces with the largest number of levels, which were dramatically reduced during the hurricane and had limited recovery thereafter. Similarly, certain low-income mixed-resident groups also have reduced hierarchies in their activity spaces due to the weather disruption. In contrast, for younger populations and those with stable home-work structures, the hierarchy of their activity spaces either remained stable or was weakened temporarily followed by a rapid recovery. Overall, this study offers a novel methodological approach to understanding the dynamics of hierarchical structures of human activity spaces across resident groups on different locations and over time. In particular, it highlights the structural changes of activity space in response to external disruptions, such as natural disasters or major events.

      The results of this study provide key insights for sustainable urban planning. Without recognizing the hierarchical organization of residents' activity spaces, urban planners and policymakers may overlook the structural constraints and their variations in people's access to activities across social groups and in response to extreme weather disruptions, which is crucial for promoting equity and fairness in urban design and planning. There are several directions for future research. First, the hierarchical structures and travel choices in our study were based on a single travel mode, as we used taxi data in the case study to analyze travel patterns. This source does not represent mobilities of all urban residents, especially lower-income groups who cannot afford taxis as a primary mode. Future research should incorporate multiple travel modes to examine how people organize their hierarchical activity spaces through different transportation options. Additionally, while this study focused on the short- to medium-term effects of external shocks, further research is needed to explore the long-term stability and adaptability of travel hierarchies in response to changing social, economic, and environmental conditions, such as those observed during the COVID-19 pandemic.

Huff, D.L. and Jenks, G.F., 1968. A Graphic Interpretation of the Friction of Distance in Gravity Models. *Annals of the Association of American Geographers*, 58 (4), 814–824.

Jeng, J. and Fesenmaier, D.R., 2002. Conceptualizing the Travel Decision-Making Hierarchy: A Review of Recent Developments. *Tourism Analysis*, 7 (1), 15–32.

Jeong, J., Lee, J., and Gim, T.T., 2022. Travel mode choice as a representation of travel utility: A multilevel approach reflecting the hierarchical structure of trip, individual, and neighborhood characteristics. *Papers in Regional Science*, 101 (3), 745–766.

Jin, M., Gong, L., Cao, Y., Zhang, P., Gong, Y., and Liu, Y., 2021. Identifying borders of activity spaces and quantifying border effects on intra-urban travel through spatial interaction network. *Computers, Environment and Urban Systems*, 87, 101625.

Kar, A., Le, H.T.K., and Miller, H.J., 2023. Inclusive Accessibility: Integrating Heterogeneous User Mobility Perceptions into Space-Time Prisms. *Annals of the American Association of Geographers*, 113 (10), 2456–2479.

Kar, A., Xiao, N., Miller, H.J., and Le, H.T.K., 2024. Inclusive accessibility: Analyzing socio-economic disparities in perceived accessibility. *Computers, Environment and Urban Systems*, 114, 102202.

Koylu, C., Tian, G., and Windsor, M., 2023. Flowmapper.org: a web-based framework for designing origin–destination flow maps. *Journal of Maps*, 19 (1), 1996479.

Kwan, M.-P., 1998. Space-Time and Integral Measures of Individual Accessibility: A Comparative Analysis Using a Point-based Framework. *Geographical Analysis*, 30 (3), 191–216.

Kwan, M.-P. and Schwanen, T., 2016. Geographies of mobility. *Annals of the American Association of Geographers*, 106 (2), 243–256.

Lee, J.H., Davis, A.W., Yoon, S.Y., and Goulias, K.G., 2016. Activity space estimation with longitudinal observations of social media data. *Transportation*, 43 (6), 955–977.

Li, J., Zhao, P., Zhang, M., Deng, Y., Liu, Q., Cui, Y., Gong, Z., Liu, J., and Tan, W., 2025. Exploring collective activity space and its spatial heterogeneity using mobile phone signaling Data: A case of Shenzhen, China. *Travel Behaviour and Society*, 38, 100920.

**Appendix A. Validation of the spatio-temporally constrained random walk method**

To validate our proposed spatio-temporally constrained random walk method, we compared the spatial distribution of aggregated trips (visited locations) generated by our model with the original OD flow data aggregated by destination. We used Pearson correlation coefficients (PCCs) to quantify the similarity between the two distributions, as presented in Table A1.

Table A1. PCCs between generated and observed spatial distributions of visited locations across different phases.

|         | Pre-event | During-event | Post-event | Overall |
|---------|-----------|--------------|------------|---------|
| PCC     | 0.8700    | 0.8888       | 0.8603     | 0.8648  |
| p-value | <0.0001   | <0.0001      | <0.0001    | <0.0001 |

The results indicate a consistently high correlation across the three different phases, demonstrating that the generated visited locations align well with the observed travel patterns. Notably, during the hurricane event, the similarity was highest, suggesting that our method remains robust even under external disruptions. This robustness may be attributed to travel behavior becoming more constrained during such events, with movements concentrated around specific locations. Our approach can effectively capture these changes, demonstrating its reliability in modeling hierarchical travel patterns across different mobility conditions.

**Appendix B. Changes in the hierarchy of taxi travel before, during, and after Hurricane Sandy: a compression effect**

We used HARM to construct hierarchical structures of taxi travels across different stages of an external shock (Hurricane Sandy)—before, during, and after the event itself. Through the lens of changes in movement properties, we can assess its impact on the hierarchical structures of mobility within functional spaces. These effects can be quantitatively expressed through several model properties, such as the number of scales (or levels), the level distance, and the size of each scale. These are considered consecutively hereafter.

*Number of levels.* Within a certain study area, a larger number of levels is indicative of a broader and more diverse range of spatial choices for activities, while fewer levels suggest a more concentrated selection of travel spaces. Figure B1 shows how the number of levels varies within the hierarchical space using the HARM-Euclidean. Since we fitted the model to each spatial unit individually, there are three phases, with 236 different hierarchies in each phase. Figure B1a depicts the overall frequency distribution of the number of levels by phase. It is apparent that during the event, the number of levels decreased compared to the pre-event phase, leading to what we term 'scale compression'. This is manifested by a decrease in the frequency of higher levels and an increase in the frequency of lower levels (since the total frequency is fixed, which corresponds to the number of spatial units). In terms of actual travel behavior, extreme weather conditions lead people to complete only essential trips, which can cause certain characteristic spatial scales to 'disappear.' After the event, the number of levels rose again, even higher than the pre-event number.

Figure B1. Distribution of number of levels and changes before, during and after the event within the hierarchical space using the HARM-Euclidean (N = 236 census tracts).

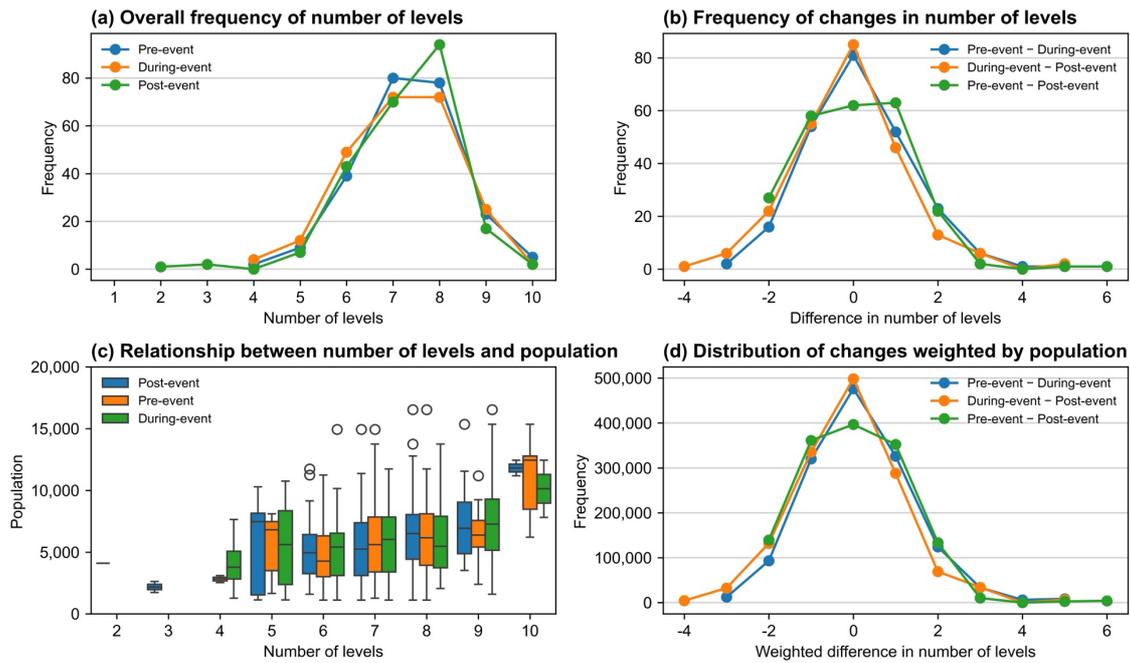

Figure B1b provides a depiction of how the number of levels in each spatial unit varies at different stages of the Hurricane event. For instance, let us consider the case of a spatial unit where the number of levels changed from 7 to 6 to 8 across the pre-event, during-event, and post-event phases; the first change value is calculated as 7−6=1, indicating a reduction in travel levels due to the event. The second change value is 6−8=−2, showing a post-event recovery with increased travel levels. The third change value compares the number of levels during pre- and post-event phases, resulting in 7−8=−1, which suggests that the hierarchical structure of travel becomes even stronger after the event. The distribution of these change values generally centers around zero. The distribution of the first difference skews to the right, indicating a reduction in travel levels due to the event, which is then countered by a recovery post-event (as evidenced by a left-skewed distribution). Similar to the overall frequency distribution, the data reveal that post-event travel activities exhibit a stronger hierarchical structure than before the event.

Furthermore, we examined the relationship between the number of levels and the population at different locations, as shown in Figure B1c. A positive correlation is observed between population count and the number of levels. On the one hand, simulating trajectories iteratively weighted by population at the same location allows areas with larger populations to capture a broader range of travel choices. However, this effect is not direct; the actual hierarchical levels are also influenced by the location itself and by the spatial metric used. Based on this, Figure B1d shows the distribution of changes weighted by population. Unlike the original finding that travel activity hierarchies are stronger post-event than pre-event, this weighted result suggests a stability between the two, with similar total changes despite some fluctuations.

By expanding the spatial metrics beyond the Euclidean distance, we analyzed the changes in the number of levels within the travel distance space and travel time space, as shown in Figures B2-B3. Across both functional spaces, the number of levels exhibits a similar pattern of 'scale compression' during the event as observed in the original space by physical distance, followed by a corresponding degree of recovery after the Hurricane. However, compared to the Euclidean space (HARM-Euclidean), the hierarchies within the travel relation spaces are more complex, with a frequency distribution that is more evenly spread. The changes in these spaces are also less likely to follow the same single-peak pattern seen in the Euclidean space.

Figure B2. Distribution of number of levels and changes before, during and after the event within the functional space of travel distance.

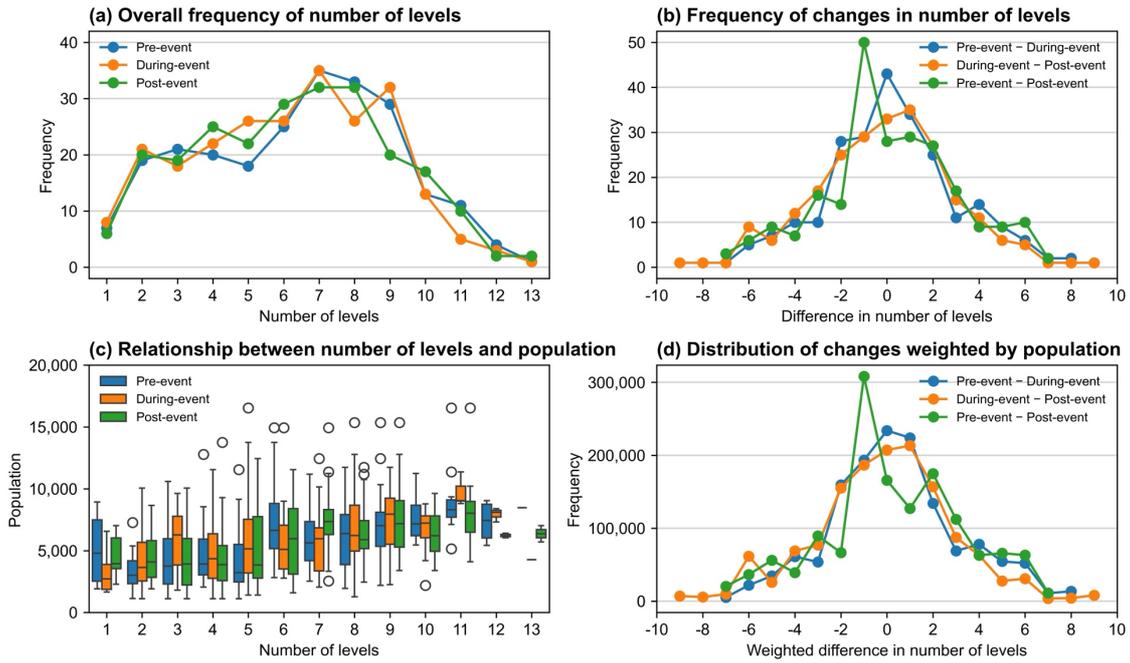

Figure B3. Distribution of number of levels and changes before, during and after the event within the functional space of travel time.

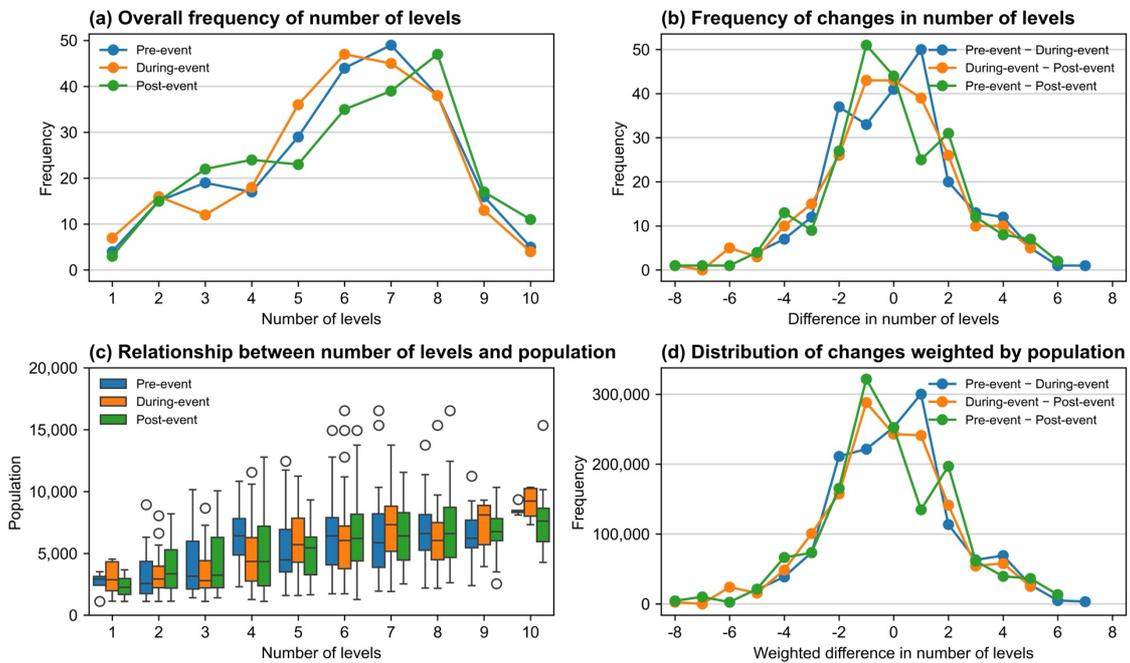

*Level distance.* Another important property is level distance, which is based on the hierarchical structure and represents the largest characteristic scale traversed during a single trip. For instance, a trip within the same neighborhood has a relatively small level distance, while travel across districts would exhibit a larger level distance. The magnitude of the level distance is also related to the number of levels, as its maximum value corresponds to the highest level in the hierarchy. As shown in Figure B4, overall, the level distance exhibits similar patterns to the number of levels, with a decrease in higher-level distances and an increase in medium and lower-level distances during the event.

Figure B4. Frequency of level distance before, during and after the event.

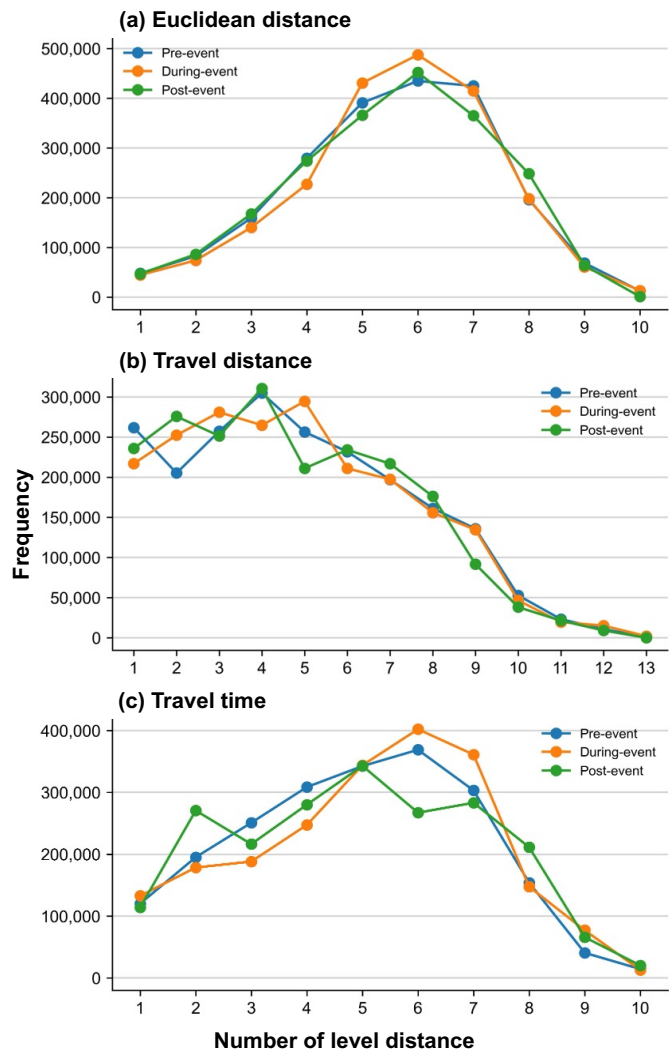

***Size of level.*** As previously mentioned, HARM allows us to identify different characteristic spatial scales of travel. Hence, we used the size of each level within the corresponding space. Specifically, this size refers to the maximum distance between any two locations within the same activity region at a given level. The size of level is derived from the optimization that generates the hierarchical structure via the complete linkage algorithm. We examined the distribution of sizes of level under different spatial measures, as shown in Figure B5. By observing changes before, during, and after the event, we can assess the impact of external shocks on the hierarchical structure of travel.

Figure B5. Distribution of sizes of level under different spatial measures before, during and after the event.

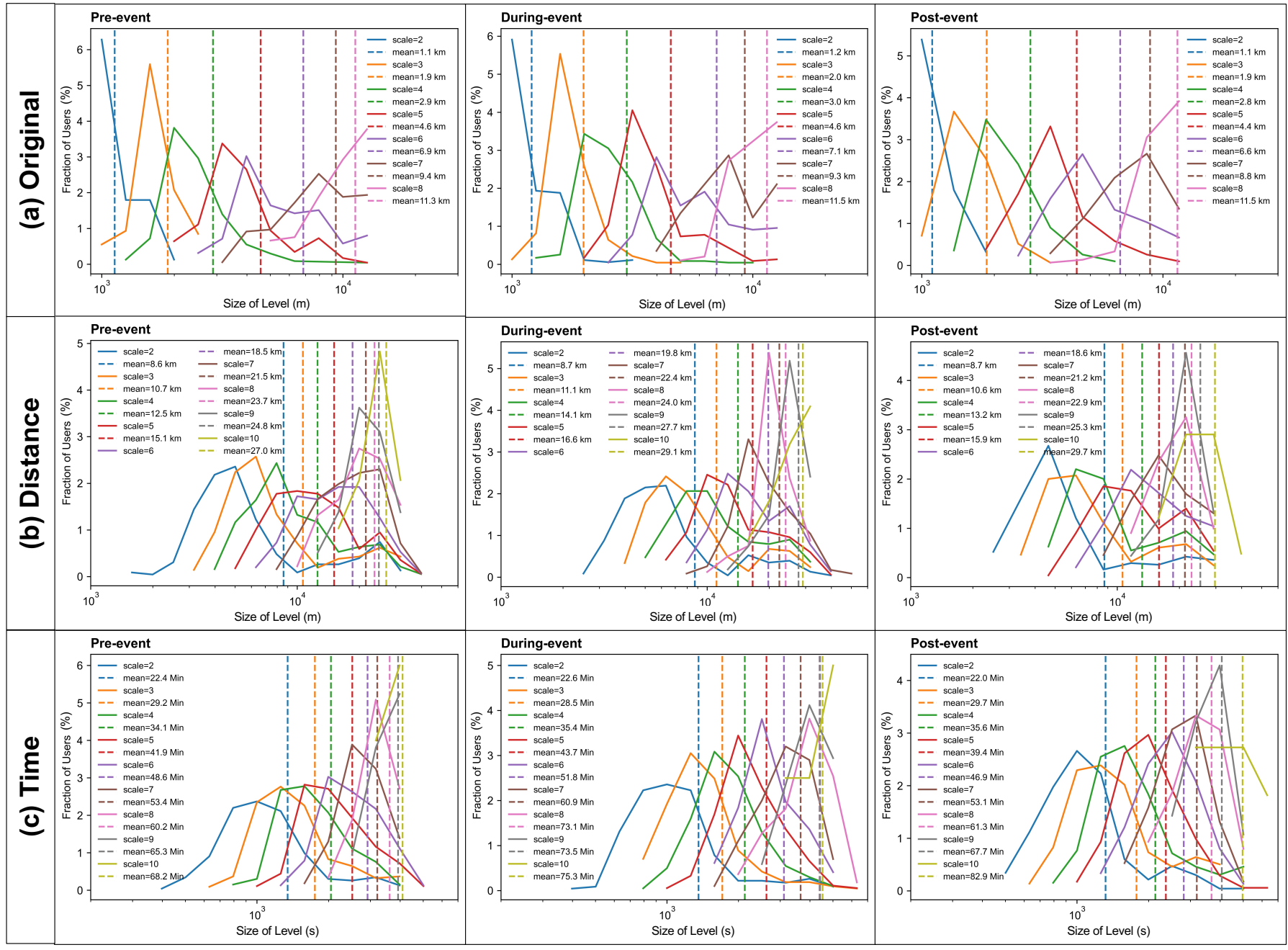

From Figure B5a, we can see that in the Euclidean space, the characteristic spatial scales show a slight increase in size during the event compared to the pre-event phase, followed by a small decrease post-event, exhibiting rather small overall changes across the three phases. However, for the results of HARM-functional, these changes were more pronounced (see Figure B5b and B5c), i.e., during the event, the sizes of each level increased significantly. For instance, in the travel distance results, at scale 4, the overall distribution shifts rightward, with the median increasing from 12.5 km to 14.1 km, and then decreasing to 13.2 km post-event. In line with the previous findings that the number of levels decreases, we observe an overall compression of spatial hierarchy denoted here by fewer levels being encapsulated within spaces of the same size.